\documentclass[aps,twocolumn,showpacs,superscriptaddress,pr]{revtex4}

\usepackage{amssymb}
\usepackage{amsmath}
\usepackage{colordvi}
\usepackage{amsthm}
\usepackage{epsfig}
\usepackage{graphicx}
\usepackage{subfigure}
\usepackage{caption}

\def\avg#1{\langle#1\rangle}
\def\Re{\rm{Re}}
\def\Im{\rm{Im}}
\def\be{\begin{equation}}
\def\ee{\end{equation}}
\def\bea{\begin{eqnarray}}
\def\eea{\end{eqnarray}}

\def\nn{\nonumber}

\def\Re{\mbox{Re}}

\begin{document}
\title{3D quaternionic condensations, Hopf invariants, and skyrmion
lattices with synthetic spin-orbit coupling}
\author{Yi Li}
\affiliation{Department of Physics, University of California, San Diego,
La Jolla, California 92093, USA}
\affiliation{Princeton Center for Theoretical Science, 
Princeton University, Princeton, NJ 08544}
\author{ Xiangfa Zhou}
\affiliation{
Key Laboratory of Quantum Information, University of Science and Technology
of China, CAS, Hefei, Anhui 230026, China
}
\author{Congjun Wu}
\affiliation{Department of Physics, University of California, San Diego,
La Jolla, California 92093, USA}

\begin{abstract}
We study the topological configurations of the two-component condensates
of bosons
with the $3$D $\vec{\sigma}\cdot \vec{p}$ Weyl-type spin-orbit coupling
subject to a harmonic trapping potential.
The topology of the condensate wavefunctions manifests in the quaternionic 
representation.
In comparison to the $U(1)$ complex phase, the quaternionic phase manifold 
is $S^3$ and the spin orientations form the $S^2$ Bloch sphere
through the 1st Hopf mapping. 
The spatial distributions of the quaternionic phases exhibit the 3D 
skyrmion configurations, and the spin distributions
possess non-trivial Hopf invariants.
Spin textures evolve from the concentric distributions at the weak 
spin-orbit coupling regime to the rotation symmetry breaking patterns
at the intermediate spin-orbit coupling regime. 
In the strong spin-orbit coupling regime, the single-particle spectra
exhibit the Landau-level type quantization. 
In this regime, the three-dimensional skyrmion lattice structures are 
formed when interactions are below the energy scale of Landau level
mixings. 
Sufficiently strong interactions can change condensates into spin-polarized
plane-wave states, or, superpositions of two plane-waves exhibiting
helical spin spirals. 
\end{abstract}
\pacs{03.75.Mn, 03.75.Lm, 03.75.Nt, 67.85.Fg }
\maketitle

\section{Introduction}
Quantum mechanical wavefunctions generally speaking are complex-valued.
However, for the single component boson systems, their ground state 
many-body wavefunctions are highly constrained, which are
usually positive-definite \cite{feynman1972}, 
as a consequence of the Perron-Frobenius theorem in the mathematical 
context of matrix analysis \cite{bapat1997}.
This is a generalization of the ``no-node'' theorem of the single-particle
quantum mechanics, for example, both the ground state wavefunctions
of harmonic oscillators and hydrogen atoms are nodeless.
Although the positive-definiteness does not apply to the many-body
fermion wavefunctions because Fermi statistics necessarily leads to 
nodal structures, it remains valid for many-body boson systems. 
It applies under the following conditions: the Laplacian type
kinetic energy, the arbitrary single-particle potential, and 
the coordinate-dependent interactions.
The positive-definiteness of the ground state wavefunctions implies that 
time-reversal (TR) symmetry cannot be spontaneously broken in 
conventional Bose-Einstein condensates (BEC), such as the 
superfluid $^4$He and most ground state BECs of ultra-cold 
alkali bosons \cite{leggett2001}.

It would be interesting to seek unconventional BECs beyond the constraint 
of positive-definite condensate wavefunctions \cite{wu2009}.
The spin-orbit coupled boson systems are an ideal platform to 
study this class of exotic states of bosons, which can spontaneously 
breaking the TR symmetry.
In addition to a simple Laplacian, the kinetic energy contains
the spin-orbit coupling term linearly dependent on momentum.
If the bare interaction is spin-independent, the condensate wavefunctions
are heavily degenerate. 
An ``order-from-disorder'' calculation based
on the zero-point energy of the Bogoliubov spectra was perform to
select the condensate configuration \cite{wu2009}.
Inside the harmonic trap, it is predicted that the condensates spontaneously 
develop the half-quantum vortex coexisting with 2D skyrmion-type 
spin textures \cite{wu2008}.
Experimentally, spin-orbit coupled bosons have been realized 
in exciton systems in semi-conducting quantum wells. 
Spin texture configurations similar to those predicted in Ref. \cite{wu2008}
have been observed \cite{high2012}.
On the other hand, the progress of synthetic artificial gauge fields 
in ultracold atomic gases greatly stimulates the investigation of the
above exotic states of bosons \cite{lin2009,lin2011}.
Extensive studies have been performed for bosons with the 2D Rashba spin-orbit
coupling, which exhibit various spin structures arising from the competitions
among the spin-orbit coupling, interaction, and the confining trap energy
\cite{stanescu2008,wu2008,ho2011,wang2010,yip2011,zhang2012,zhou2011,hu2012,
santos2011}.

Most studies so far have been on the two-dimensional spin-orbit 
coupled bosons. 
It would be interesting to further consider the unconventional
condensates of bosons with the three-dimensional Weyl-type 
spin-orbit coupling, whose experimental realization has been
proposed by the authors through atom-light interactions in a combined 
tripod and tetrapod level system \cite{li2012} and also by Anderson 
{\it et al.} \cite{anderson2011}.
As will be shown below, the quaterinon representation provides
a natural and most beautiful description of the topological
condensation configurations. 
Quaternions are an extension of complex numbers as the first discovered
non-commutative division algebra, which has provided a new formulation 
of quantum mechanics \cite{adler1995,finkelstein1962,balatsky1992}.
Similarly to complex numbers whose phases span a unit circle $S^1$, the
quaternionic phases span a three dimensional unit sphere $S^3$.
The spin distributions associated with quaternionic
wavefunctions are obtained through the 1st Hopf map $S^3\rightarrow S^2$
as will be explained below.
It would be interesting to search for BECs with non-trivial
topological defects associated with the quaternionic phase structure.
It will be a new class of unconventional BECs beyond the
``no-node'' theorem breaking TR symmetry spontaneously.

In this article, we consider the unconventional condensate 
wavefunctions with the 3D Weyl-type spin-orbit coupling 
$\vec \sigma \cdot \vec p$.
The condensation wavefunctions exhibit topologically non-trivial 
configurations as 3D skyrmions, and spin density distributions are 
also non-trivial with non-zero Hopf invariants.
These topological configurations can be best represented 
as defects of quaternion phase distributions.
Spatial distributions of the quaternionic phase textures and spin textures
are concentric at weak spin-orbit couplings.
As increasing spin-orbit coupling, these textures evolve to lattice structures
which are the 3D quaternionic analogy of the 2D Abrikosov lattice
of the usual complex condensate.

The rest part of this article is organized as follows.
In Sect. \ref{sect:model}, we define the model Hamiltonian.
In Sect. \ref{sect:weak}, the condensate wavefunctions in the weak 
spin-orbit regime are studied.
Topological analyses on the skyrmion configurations and
Hopf invariants are performed by using the quaternion representation.
In Sect. \ref{sect:inter}, the skyrmion lattice configuration
of the spin textures is studied in the intermediate and strong
spin-orbit coupling regimes.
In Sect. \ref{sect:strong}, superpositions of plane-wave
condensate configurations are studied.
Conclusions are made in Sect. \ref{sect:conc}.

%%%%%%%%%%%%%%%%%%%%%%%%%%%%%%%%%%%%%%%%%%%%%%%%%%%%%%%%%%%%%%%%%%%%%%%%
\section{The Model Hamiltonian}
\label{sect:model}

We consider a two-component boson system with the 3D spin-orbit coupling
of the $\vec\sigma\cdot \vec p$-type confined in a harmonic trap.
The free part of the Hamiltonian is defined as
\bea
H_0& =& \int d^3 \vec{r} ~\psi_{\gamma}^{\dag}(\vec r)
\Big\{-\frac{\hbar^2\vec \nabla^2}{2 m}
+i\hbar\lambda \vec{\sigma}_{\gamma\delta} \cdot (\vec \nabla) \nn \\
&+&
\frac{1}{2}m\omega^2 \vec{r}\,^2 \Big\}  \psi_\delta (\vec r),
\label{eq:ham_0}
\eea
where $\gamma$ and $\delta$ equal $\uparrow$ and $\downarrow$ 
referring to two internal states of 
bosons; $\vec \sigma$ are Pauli matrices; $m$ is the boson mass;
$\lambda$ is the spin-orbit coupling strength with the unit 
of velocity; $\omega$ is the trap frequency.
At the single-particle level, Eq. (\ref{eq:ham_0}) satisfies the Kramer-type
time-reversal symmetry of $T=(-i\sigma_2) C$ with the property of $T^2=-1$.
However, parity is broken by spin-orbit coupling.
In the absence of the trap, good quantum numbers for the single-particle
states are the eigenvalues $\pm 1$ of helicity $\vec \sigma \cdot \vec p/|p|$,
where $p$ is the momentum.
This results in two branches of dispersions
\bea
\epsilon_\pm (\vec k)=\frac{\hbar^2}{2m}(k \mp k_{so})^2,
\eea
where $\hbar k_{so}=m\lambda$.
The lowest single-particle energy states lie in the sphere with the radius
$k_{so}$ denoted as the spin-orbit sphere.
It corresponds to a spin-orbit length scale $l_{so}=1/k_{so}$ in real space.
The harmonic trap has a natural length scale
$l_{T}=\sqrt{\frac{\hbar}{m\omega}}$, and thus
the dimensionless parameter $\alpha=l_T k_{so}$ describes
the relative spin-orbit coupling strength.

As for the interaction Hamiltonian, we use the contact $s$-wave 
scattering interaction defined as
\bea
H_{int} =\frac{g_{\gamma\delta}}{2}\int d^3 \vec{r} ~
\psi_{\gamma}^{\dag}(\vec{r})
\psi_{\delta}^{\dag}(\vec{r})
\psi_{\delta}(\vec{r}) \psi_{\gamma}(\vec{r}).
\eea
Two different interaction parameters are allowed, including the intra and
inter-component ones defined as $g_{\uparrow\uparrow} =g_{\downarrow\downarrow}
= g$, and $g_{\uparrow\downarrow}=c g$, where $c$ is a constant.

In the previous study of the 2D Rashba spin-orbit coupling with harmonic 
potentials \cite{wu2008,hu2012}, the single-particle eigenstates are 
intuitively expressed in the momentum representation: the low energy
state lies around a ring in momentum space, and the harmonic potential
becomes the planar rotor operator on this ring subject to a $\pi$-flux,
which quantizes the angular momentum $j_z$ to half integers.
Similar picture also applies in 3D \cite{ghosh2011,wu2008}.
The low energy states are around the spin-orbit sphere.
In the projected low energy Hilbert space, the eigenvectors read
\bea
\psi_+(\vec k)= (\cos\frac{\theta_k}{2},
\sin\frac{\theta_k}{2}e^{i\phi_k})^T.
\eea
The harmonic potential is again a rotor Hamiltonian on the spin-orbit sphere
subject to the Berry gauge connection as
\bea
V_{tp}=\frac{1}{2}m (i \nabla_k -\vec A_k)^2
\eea
with the moment of
inertial $I=M_k k_{so}^2$ and $M_k=\hbar^2/(m\omega^2)$.
$\vec A_k=i\avg{\psi_{+}(\vec k)|\nabla_k|\psi_{+}(\vec k)}$ is the vector
potential of a $U(1)$ magnetic monopole, which quantizes the
angular momentum $j$ to half-integers.
While the radial energy is still quantized in terms of $\hbar \omega$,
the angular energy dispersion with respect to $j$ is strongly
suppressed at large values of $\alpha$ as
\bea
E_{n_r,j,j_z}\approx \Big(n_r
+\frac{j(j+1)}{2\alpha^2} \Big ) \hbar \omega +\mbox{const},
\label{eq:3D_SO_gap}
\eea
where $n_r$ is the radial quantum number.
As further shown in Ref. \cite{li2012}, in the case $\alpha\gg 1$, all the
states with the same $n_r$ but different $j$ and $j_z$ are nearly degenerate,
thus can be viewed as one 3D Landau level with spherical symmetry but 
the broken parity.
If filled with fermions, the system belongs to the $Z_2$-class of
3D strong topological insulators.

Now we load the system with bosons.
The interaction energy scale is defined as $E_{int}= g
N_0/l_T^3$, where $N_0$ is the total particle number in the condensate.
The corresponding dimensionless parameter is $\beta=E_{int}/\hbar \omega$.
At the Hartree-Fock level, the Gross-Pitaevskii energy functional is
defined in terms of the condensate wavefunction
$\Psi=(\Psi_\uparrow,\Psi_\downarrow)^T$ as
\bea
E&=&\int d^3 \vec r ~(\Psi_\uparrow^\dagger,\Psi_\downarrow^\dagger)
\Big\{-\frac{\hbar^2\nabla^2}{2m}
-i\lambda\hbar \vec \nabla \cdot \vec \sigma
+\frac{1}{2}m\omega^2 r^2\nn \\
&+&
 g \left ( \begin{array}{cc} n_\uparrow + c n_\downarrow & 0 \\
0 & c n_\uparrow + n_\downarrow
\end{array}
\right )  \Big\}
\left ( \begin{array}{c} \Psi_{\uparrow} \\
\Psi_{\downarrow}
\end{array}
\right ),
\label{eq:GP}
\eea
where $n_{\uparrow,\downarrow}(\vec r)=N_0|\Psi_{\uparrow,\downarrow}(\vec r)|^2$
are the particle densities of two components, respectively,
and $\Psi(\vec r)$ is normalized as
$\int d^3 \vec r \Psi^\dagger (\vec r) \Psi(\vec r)=1$.
The condensate wavefunction $\Psi(\vec r)$ is solved numerically by 
using the standard method of imaginary time evolution.
The dimensionless form of the Gross-Pitaevskii equation is
\bea
E^\prime &=&\int d^3 \vec r^\prime ~(\tilde{\Psi}_\uparrow^\dagger,
\tilde{\Psi}_\downarrow^\dagger)
\Big\{
-\frac{\vec \nabla^{\prime 2}}{2} 
-i\alpha \vec \nabla^\prime \cdot \vec \sigma
+\frac{r^{\prime 2}}{2} \nn \\
&+&
\beta \left ( \begin{array}{cc} \tilde{n}_\uparrow + c \tilde{n}_\downarrow & 0 \\
0 & c \tilde{n}_\uparrow + \tilde{n}_\downarrow
\end{array}
\right )  \Big\}
\left ( \begin{array}{c} \tilde{\Psi}_{\uparrow} \\
\tilde{\Psi}_{\downarrow}
\end{array}
\right ),
\eea
where $E^\prime=E/(\hbar\omega)$,
$\vec \nabla^\prime=l_T\vec \nabla$; $\vec r^\prime=\vec r/l_T$;
$\tilde{\Psi}_\uparrow$ and $\tilde{\Psi}_\downarrow$ are the 
renormalized condensate wavefunctions satisfying
$\int d^3 r^\prime |\tilde{\Psi}_\uparrow|^2 + |\tilde{\Psi}_\downarrow|^2=1$;
$\tilde n_\uparrow= |\tilde{\Psi}_\uparrow|^2$
and $\tilde n_\downarrow= |\tilde{\Psi}_\downarrow|^2$.

%%%%%%%%%%%%%%%%%%%%%%%%%%%%%%%%%%%%%%%%%%%%%%%%%%%%%%%%%
\section{The weak spin-orbit coupling regime}
\label{sect:weak}

In this section, we consider the condensate configuration in the
limit of weak spin-orbit coupling, say, $\alpha\sim 1$.
In this regime, the single-particle spectra still resemble
those of the harmonic trap. 
We study the case that interactions are not strong enough
to mix states with different angular momenta.

\subsection{The spin-orbit coupled condensate}
In this regime, the condensate wavefunction $\Psi$ remains the same 
symmetry structure as the single-particle wavefunction over a wide 
range of interaction parameter  $\beta$, i.e.,
$\Psi$ remains the eigenstates of $j=\frac{1}{2}$ as confirmed numerically
below.
$\Psi$ can be represented as
\bea
\Psi_{j=j_z=\frac{1}{2}}(r,\hat\Omega) &=&
f(r) Y^+_{j,j_z}(\hat\Omega)
+ig(r) Y^-_{j,j_z} (\hat\Omega), \ \ \, \ \ \,
\label{eq:cond_1}
\eea
where $f(r)$ and $g(r)$ are real radial functions.
$Y^\pm_{j,j_z}(\hat\Omega)$ are the spin-orbit coupled spherical harmonic 
functions with even and odd parities, respectively.
For example, for the case of $j=j_z=\frac{1}{2}$, they are
\bea
Y^+_{\frac{1}{2},\frac{1}{2}}(r,\hat\Omega)=\left(
\begin{array}{c}
1\\
0
\end{array}
\right), \ \ \,
Y^-_{\frac{1}{2},\frac{1}{2}}(r,\hat \Omega)=\left(
\begin{array}{c}
\cos\theta\\
\sin\theta e^{i\phi}
\end{array}
\right),
\eea
whose orbital partial-wave components 
are $s$ and $p$-wave, respectively. 
The TR partner of Eq. (\ref{eq:cond_1}) is $\psi_{j_z=-\frac{1}{2}}
= \hat{T} \psi_{j=j_z=\frac{1}{2}}= i\sigma_2 \psi_{j=j_z=\frac{1}{2}}^*$.
The two terms in Eq. (\ref{eq:cond_1}) are of opposite parity eigenvalues,
mixed by the parity breaking spin-orbit coupling 
$\vec \sigma \cdot \vec p$.
The coefficient $i$ of the $Y^-_{jj_z}$ term is because
the matrix element $\avg{Y^+_{jj_z}|\vec \sigma \cdot \vec p |Y^-_{jj_z}}$
is purely imaginary.

For the non-interacting case, the radial wavefunctions up to a Gaussian
factor can be approximated by spherical Bessel functions as
\bea
f(r)\approx j_0(k_{so}r)e^{-r^2/2l_T^2}, \ \ \,
g(r)\approx j_1(k_{so}r) e^{-r^2/2l_T^2},
\eea
which correspond to the $s$ and $p$-partial waves, respectively.
Both of them oscillate along the radial direction and the
pitch values are around $k_{so}$. 
At $r=0$, $f(r)$ reaches the maximum and $g(r)$ is 0.
As $r$ increases, roughly speaking, the zero points of $f(r)$
corresponds to the extrema of $g(r)$ and vise versa.
Repulsive interactions expand the spatial distributions of $f(r)$ and
$g(r)$, but the above picture still holds qualitatively.
In other words, there is a $\frac{\pi}{2}$-phase shift between
the oscillations of $f(r)$ and $g(r)$.

%%%%%%%%%%%%%%%%%%%%%%%%%%%%%%%%%%%%%%%%%%%%%%%%%%%%%%%%%%%%%%%%%%%%%%%%%%

\subsection{The quaternion representation}
Can we have unconventional BECs with non-trivial quaternionic
condensate wavefunctions?
Actually, the topological structure of condensate wavefunction 
Eq. (\ref{eq:cond_1}) manifests clearly in the quaternion representation
as shown below. 

We define the following mapping from the complex two-component 
vector $\Psi=(\Psi_\uparrow,\Psi_\downarrow)^T$ to 
a quaternion variable through 
\bea
\xi =\xi_0 +\xi_1 i +
\xi_2 j +\xi_3 k,
\eea
where
\bea
\xi_0=\Re
\Psi_\uparrow, \xi_1=\Im \Psi_\downarrow, \xi_2=-\Re\Psi_\downarrow,
\xi_3=\Im\Psi_\uparrow.
\eea
$i,j,k$ are the imaginary units satisfying $i^2=j^2=k^2=-1$,
and the anti-commutation relation $ij=-ji=k$.
The TR transformation on $\xi$ is just $-j\xi$.

Eq. (\ref{eq:cond_1}) can be expressed in the quaternionic exponential form as
\bea
\xi_{j=j_z=\frac{1}{2}}(r, \hat \Omega)=|\xi (r)| e^{ \vec \omega(\hat\Omega) \gamma(r)}
=|\xi| (\cos \gamma + \vec \omega \sin\gamma), \ \ \,
\label{eq:qua_phase}
\eea
where
\bea
&&|\xi(r)|=[f^2(r)+g^2(r)]^{\frac{1}{2}}, \nn \\
&&\vec \omega(\hat\Omega)=
\sin\theta \cos\phi ~i +\sin\theta \sin\phi~ j + \cos\theta ~ k, \nn \\
&&\cos\gamma(r)=f(r)/|\xi(r)|,  \ \ \,
\sin\gamma(r)=g(r)/|\xi(r)|. ~~
\eea
$\omega(\hat \Omega)$ is the imaginary unit along the direction
of $\hat\Omega$ satisfying $\vec \omega^2(\hat\Omega)=-1$.
According to the oscillating properties of $f(r)$ and $g(r)$,
$\gamma(r)$ spirals as $r$ increases.
At the $n$-th zero point of $g(r)$ denoted $r_n$, $\gamma(r_n)=n\pi$
where $n\ge 0$ and we define $r_0=0$,
while at the $n$-th zero point of $f(r)$ denoted $r^\prime_n$,
$\gamma(r_n^\prime)=(n-\frac{1}{2})\pi$ where $n\ge 1$.

In 3D, the condensate wavefunctions can be topologically non-trivial
because the homotopy group of the quaternionic phase is
$\pi_3(S^3)=Z$ \cite{wilczek1983,nakahara2003}.
The corresponding winding number, i.e. the Pontryagin index,
of the mapping $S^3\rightarrow S^3$ is the 3D skyrmion number.
The spatial distribution of the quaternionic phase 
$e^{\vec\omega(\hat\Omega)\gamma(r)}$ defined in Eq. \ref{eq:qua_phase},
which lies on $S^3$, exhibits a topologically nontrivial mapping from
$R^3$ to $S^3$, i.e.,  a 3D multiple skyrmion configuration.
This type of topological defects are non-singular which is different
from the usual vortex in single component BEC.
In realistic trapping systems, the coordinate space is the open $R^3$.
At large distance $r\gg l_T$, $|\xi(r)|$ decays exponentially, where
the quaternionic phase and the mapping are not well-defined.
Nevertheless, in each concentric spherical shell with
$r_n< r <r _{n+1}$, $\gamma(r)$ winds
from $n\pi$ to $(n+1)\pi$, and $\omega(\hat \Omega)$
covers all the directions, thus this shell contributes 1 to the
winding number of $e^{\vec\omega(\hat\Omega)\gamma(r)}$ on $S^3$.
If  the system size is truncated at the order of $l_T$,
the skyrmion number can be approximated at the order of
$l_T k_{so}=\alpha$. 

There exists an interesting difference from the previously studied 
2D case: Although the spin density distribution
exhibit the 2D skyrmion configuration due to $\pi_2(S^2)$
\cite{wu2008,hu2012,santos2011}, the 2D condensation wavefunctions
have no well-defined topology due to $\pi_2(S^3)=0$.

%----------------------------------------------
\begin{widetext}
\begin{center}
\includegraphics[width=0.32\linewidth]{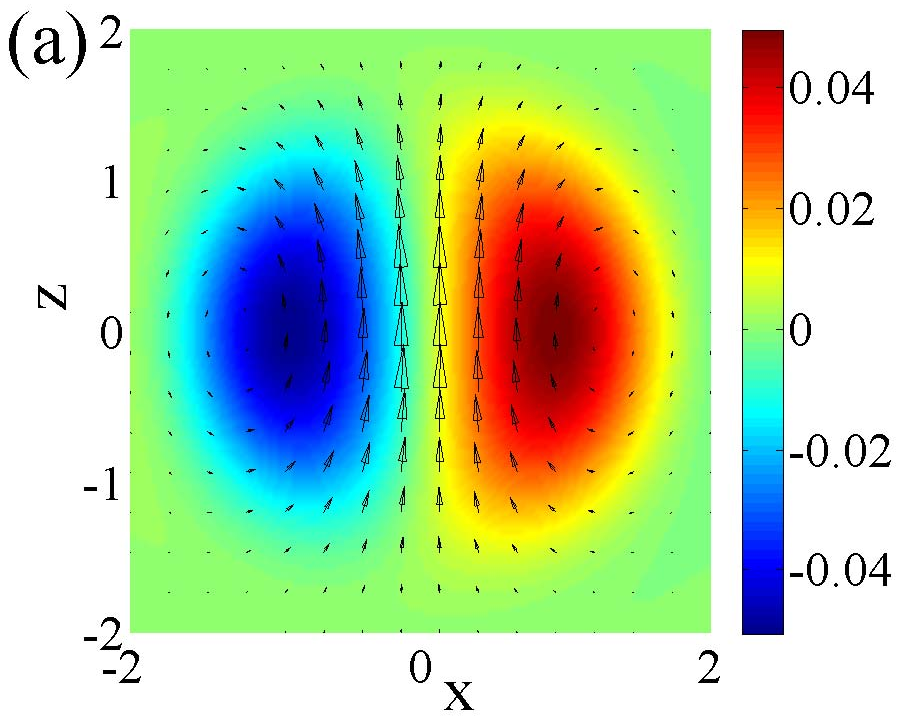}
\includegraphics[width=0.32\linewidth]{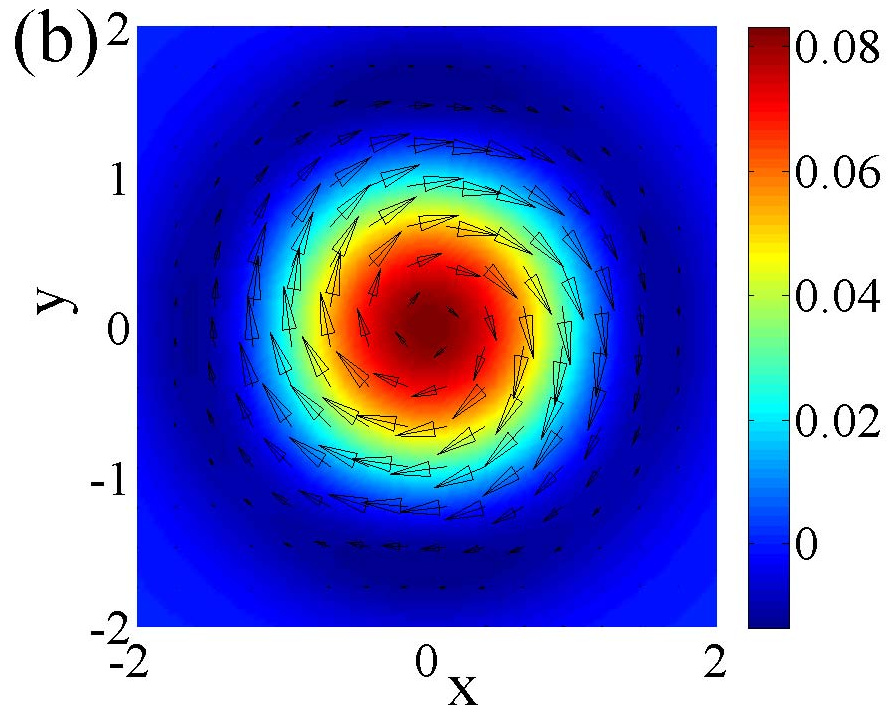}
\includegraphics[width=0.32\linewidth]{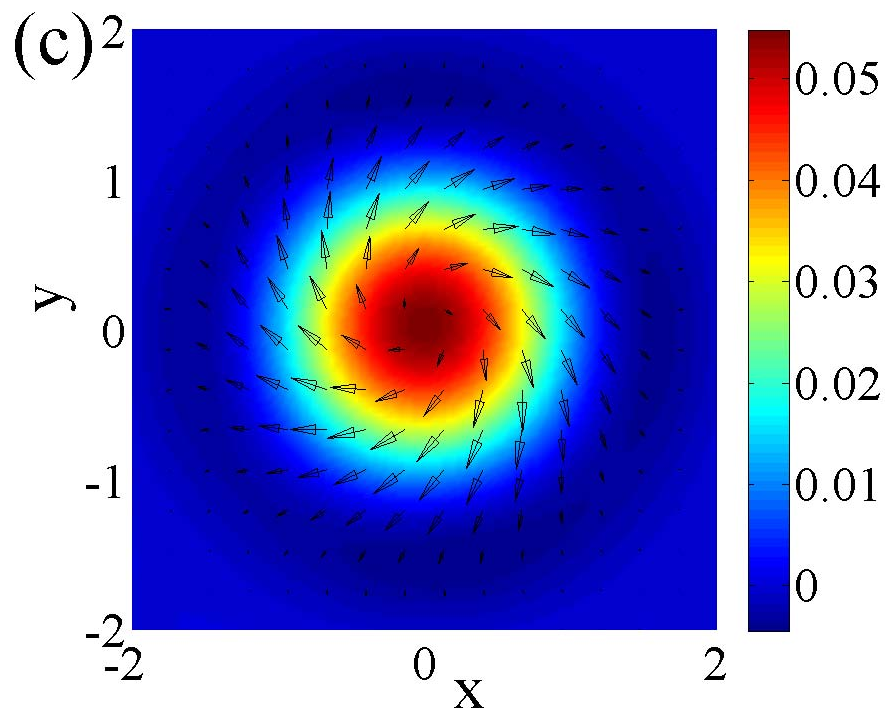}
\captionof{figure}{The distribution of $\vec{S} (\vec r)$ in a) the $xz$-plane
and in the horizontal planes with $b$) $z=0$ and $c$) $z/l_T=\frac{1}{2}$.
The unit length is set as $l_T=1$ in all the figures in this article.
The color scale shows the magnitude of out-plane component
$S_y$ in $a$) and $S_z$ in $b$) and $c$).
The parameter values are $\alpha=1.5$, $c=1$, and $\beta=30$,
and the length unit in these and all the figures below is $l_T$.
}
\label{fig:spin}
\end{center}
\end{widetext}
%------------------------------------------------------------

%-----------------------------------------------------
\subsection{The Hopf mapping and Hopf invariant}

Exotic spin textures in spinor condensates have been extensively
investigated \cite{zhou2003,zhang2009,stamper2012}.
In our case, the 3D spin density distributions $\vec S(\vec r)$ exhibit
a novel configuration with non-trivial Hopf invariants
due to the non-trivial homotopy group $\pi_3(S^2)=Z$ 
\cite{wilczek1983,nakahara2003}.
$\vec S(\vec r)$ can be obtained from $\xi(r)$ through the 1st Hopf map
defined as $\vec S (\vec r)=\frac{1}{2}\psi^\dagger_\gamma \vec\sigma_{\gamma\beta}
\psi_\delta$, or, in the quaternionic representation,
\bea
\frac{1}{2}\bar \xi k \xi=
S_x i + S_y j + S_z k,
\label{eq:hopf}
\eea
where $\bar \xi
=\xi_0 - \xi_1 i - \xi_2 j - \xi_3 k$ is the quaternionic conjugate of $\xi$.
The Hopf invariant of the 1st Hopf map is just 1
\cite{nakahara2003}.
The real space concentric spherical shell $r_{n}<r<r_{n+1}$
 maps to the
quaternionic phase $S^3$, and the latter further maps to the $S^2$ Bloch sphere
through the 1st Hopf map.
The winding number of the first map is 1, and the Hopf invariant  of the
second map is also 1, thus the Hopf invariant of the 
shell $r_{n}<r<r_{n+1}$ to $S^2$ is 1.
Rigorously speaking, the magnitude of $\vec S(\vec r)$ decays
exponentially at $r\gg l_T$, and thus the total Hopf invariant is not
well-defined in the open $R^3$ space.
Again, if we truncate the system size at $l_T$, the Hopf invariant
is approximately at the order of $\alpha$.

%------------------------------------------
\begin{center}
\includegraphics[width=0.7\linewidth]{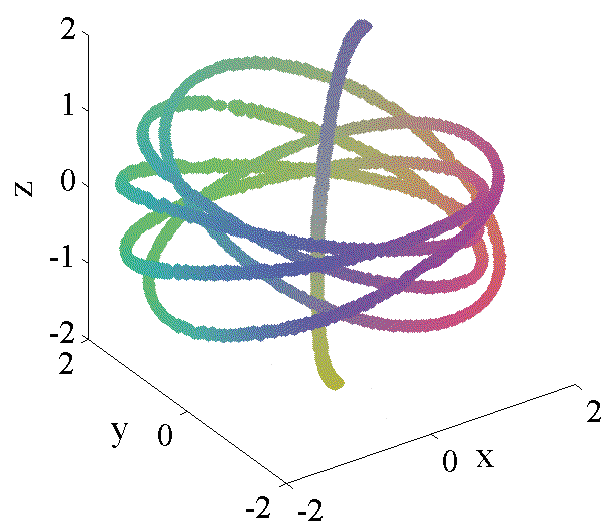}
\captionof{figure}{The Hopf fibration of the spin texture
configuration in Fig. \ref{fig:spin}.
Every circle represents a spin orientation, and every two 
circles are linked with the linking number 1.
}
\label{fig:hopf}
\end{center}
%------------------------------------------------------------

Next we present numeric results for the spin textures
associated with the condensation wavefunction Eq. \ref{eq:cond_1}
as plotted in Fig. \ref{fig:spin}.
Explicitly, $\vec S(\vec r)$ is expressed as
\bea
\left[
\begin{array}{c}
S_x(\vec r)\\
S_y(\vec r)
\end{array}
\right] &=&g(r) \sin\theta
\left[
\begin{array}{cc}
\cos\phi& -\sin\phi\\
\sin\phi& \cos\phi
\end{array}
\right] \left[
\begin{array}{c}
g(r)\cos\theta\\
f(r)
\end{array}
\right], \nn \\
S_z(\vec r)~~&=& f^2(r)+g^2(r) \cos2\theta,
\label{eq:spin}
\eea
In the $xz$-plane, the in-plane components $S_x$ and $S_z$ form a vortex
in the half plane of $x>0$ and $S_y$ is prominent in the core.
The contribution at large distance is neglected, where $\vec S(\vec r)$ decays
exponentially.
Due to the axial symmetry of $\vec S(\vec r)$ in Eq. \ref{eq:spin}, the 3D
distribution is just a rotation of that in Fig. \ref{fig:spin} a) around
the $z$-axis.
In the $xy$-plane, spin distribution exhibits a 2D skyrmion pattern, whose
in-plane components are along the tangential direction.
As the horizontal cross-section shifted along the $z$-axis,
$\vec S(\vec r)$ remains 2D skyrmion-like, but its in-plane
components are twisted around the $z$-axis.
The spin configuration at $z=-z_0$ can be obtained by a combined
operation of TR and rotation around the $y$-axis 180$^{\circ}$,
thus its in-plane components are twisted in an opposite way
compared to those at $z=z_0$.
Combining the configurations on the vertical and horizontal cross sections,
we complete the 3D distribution of $\vec S(\vec r)$ with non-zero Hopf
invariant.

The non-trivial structure of the Hopf invariant of the above spin 
configuration can be revealed by plotting its Hopf fibration in terms 
of the linked non-crossing circles in real space, as shown in 
Fig. \ref{fig:hopf}.
For all the points on each circle, their normalized spin polarizations 
$\langle \vec{\sigma} \rangle/|\langle \vec{\sigma} \rangle|$ are the same, 
corresponding to a single point on the $S^2$ sphere.
In addition, every two circles are linked with each other with the
linking number $1$, which is the standard Hopf bundle structure 
describing a many-to-one map from $S^3$  to $S^2$. 
Ultracold bosons with synthetic spin-orbit coupling provide a novel 
platform to study such beautiful mathematical ideas in realistic physics systems.

%---------------------------------------------------
\begin{widetext}
\begin{center}
\includegraphics[width=0.3\linewidth]{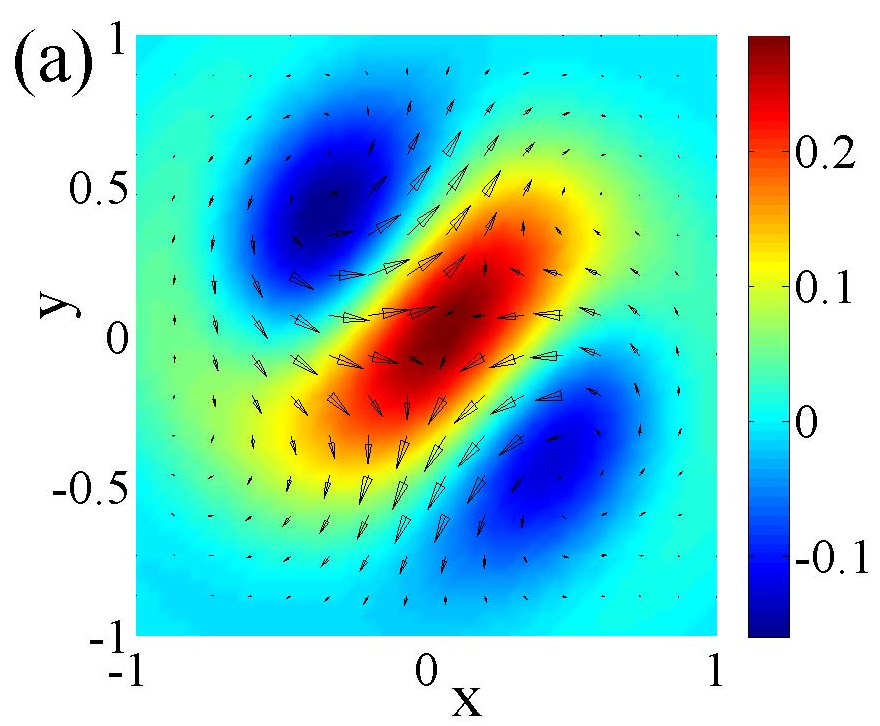} 
\includegraphics[width=0.3\linewidth]{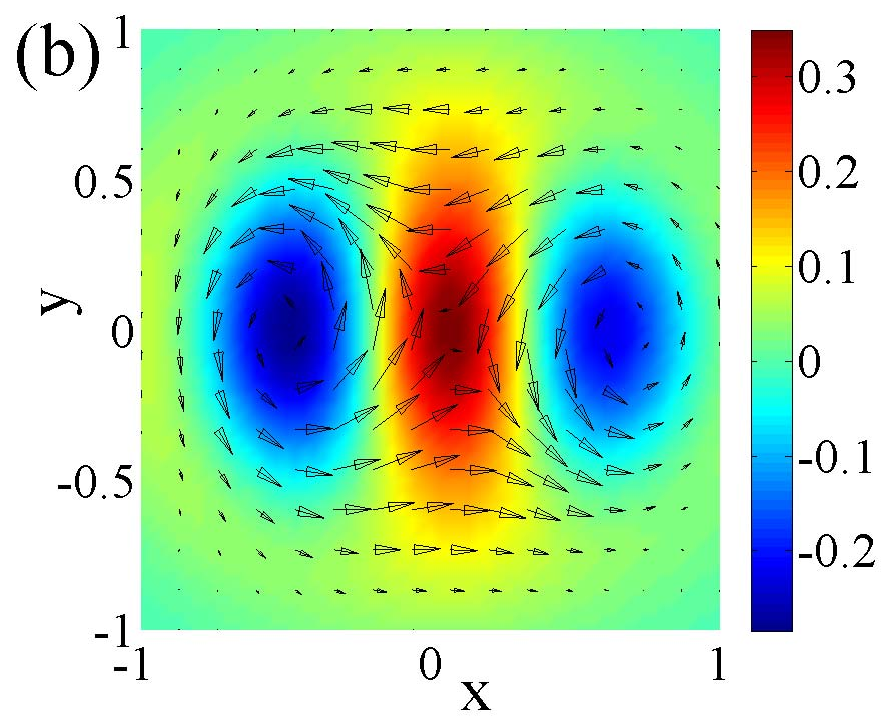}
\includegraphics[width=0.3\linewidth]{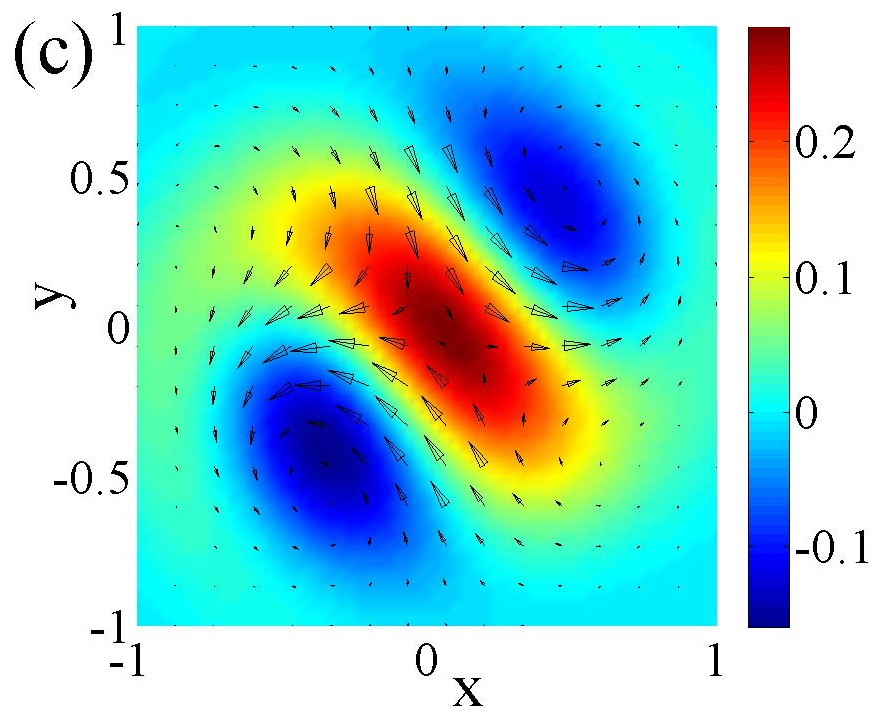}
\captionof{figure}{The distribution of $\vec S(\vec r)$ in
horizontal cross-sections with
a) $z/l_T=-0.5$, b) $z/l_T=0$, c) $z/l_T=0.5$,
respectively.
The color scale shows the value of $S_z$,
and parameter values are $\alpha=4$, $\beta=2$, and $c=1$.
}
\label{fig:spin_2}
\end{center}
\end{widetext}
%----------------------------------------------------

%%%%%%%%%%%%%%%%%%%%%%%%%%%%%%%%%%%%%%%%%%%%%%%%%%%%%%%%%%%%%%%
\section{The intermediate and strong spin-orbit coupling regime}
\label{sect:inter}

\subsection{The intermediate spin-orbit coupling strength}
Next we consider the case of the intermediate spin-orbit coupling strength, 
i.e., $1<\alpha<10$, at which the single-particle spectra evolve from 
the case of the harmonic potential to Landau level-like as shown in 
Eq. \ref{eq:3D_SO_gap}.
Interactions are sufficiently strong to mix a few lowest energy states
with different angular momenta $j$. 
As a result, rotational symmetry is broken and complex patterns appear.

In this case, the topology of condensate wavefunctions is still 3D
skyrmion-like mapping from $R^3$ to $S^3$,
and spin textures with the non-trivial Hopf invariant
are obtained through the 1st Hopf map.
Compared to the weak spin-orbit coupling case, the quaternionic phase
skyrmions and spin textures
are no longer concentric, but split to a multi-centered pattern.
The numeric results of $\vec S(\vec r)$ are plotted in Fig. \ref{fig:spin_2}
for different horizontal cross-sections.
In the $xy$-plane, $\vec S$ exhibits the 2D skyrmion pattern
as shown in Fig. \ref{fig:spin_2} ($b$):
The in-plane components form two vortices and one anti-vortex,
while $S_z$'s inside the vortex and anti-vortex cores
are opposite in direction, thus they contribute to the skyrmion number
with the same sign.
The spin configuration at $z=z_0>0$ is shown in Fig. \ref{fig:spin_2} (a),
which is twisted around the $z$-axis clock-wise.
After performing the combined TR and rotation around the $y$-axis 180
$^\circ$, we arrive at the configuration at $z=-z_0$
in Fig. \ref{fig:spin_2}(b).

%--------------------------------------------------------
\subsection{The strong spin-orbit coupling regime}
We next consider the case of strong spin-orbit coupling, i.e., $\alpha\gg 1$.
The single-particle spectra already exhibit the Landau-level type
quantization in this regime as shown in Eq. \ref{eq:3D_SO_gap}.
The single-particle eigenstates with $n_r=0$ are nearly degenerate i.e., 
they form the lowest Landau level states.
We assume that the interaction strength is enough to mix states inside
the lowest Landau level but is still relatively weak not to induce
inter-Landau level mixing. 

In this regime, the length scale of each skyrmion is shortened as 
enlarging the spin-orbit coupling strength.
As we can imagine, more and more skyrmions appear and will form a 3D 
lattice structure, which is the SU(2) generalization of the 2D Abrikosov 
lattice of the usual U(1) superfluid. 
We have numerically solved the Gross-Pitaevskii equation 
Eq. \ref{eq:GP} and found the lattice structure:
Each lattice site is a single skyrmion of the condensate 
wavefunction $\xi(\vec r)$, whose spin configuration exhibits the 
texture configuration approximately with a unit Hopf invariant.
The numeric results for the spin texture configuration are depicted 
in Fig. \ref{fig:lattice} $a$) and $b$) for two different horizontal
cross sections parallel to the $xy$-plane. 
In each cross section, spin textures form a square lattice, and 
the lattice constant $d$ is estimated approximately the
spin-orbit length scale as
\bea
d\simeq 2\pi l_{so} = 2\pi l_T/\alpha.
\eea
For two horizontal cross sections with a distance of $\Delta z\simeq
d/2$, their square lattice configurations are displaced along
the diagonal direction: The sites at one layer sit above
the plaquette centers of the adjacent layer. 
As a result, the overall three-dimensional configuration of
the topological defects is a body-centered cubic ($bcc$) lattice,
and its size is finite confined by the trap.

%----------------------------------------------------------
\begin{widetext}
\begin{center}
\includegraphics[width=0.4\linewidth]{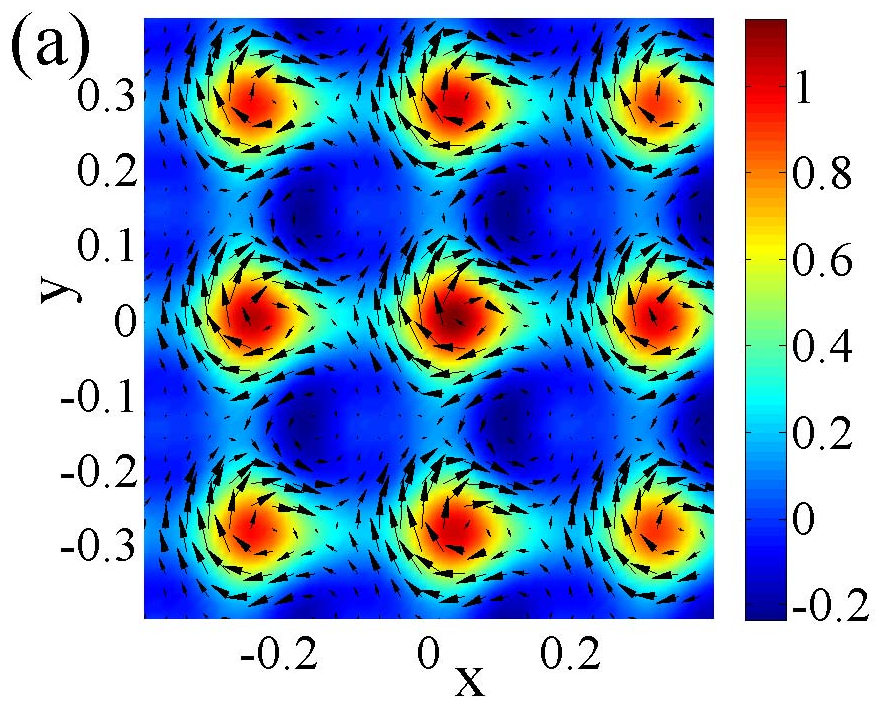} \ \ \
\includegraphics[width=0.4\linewidth]{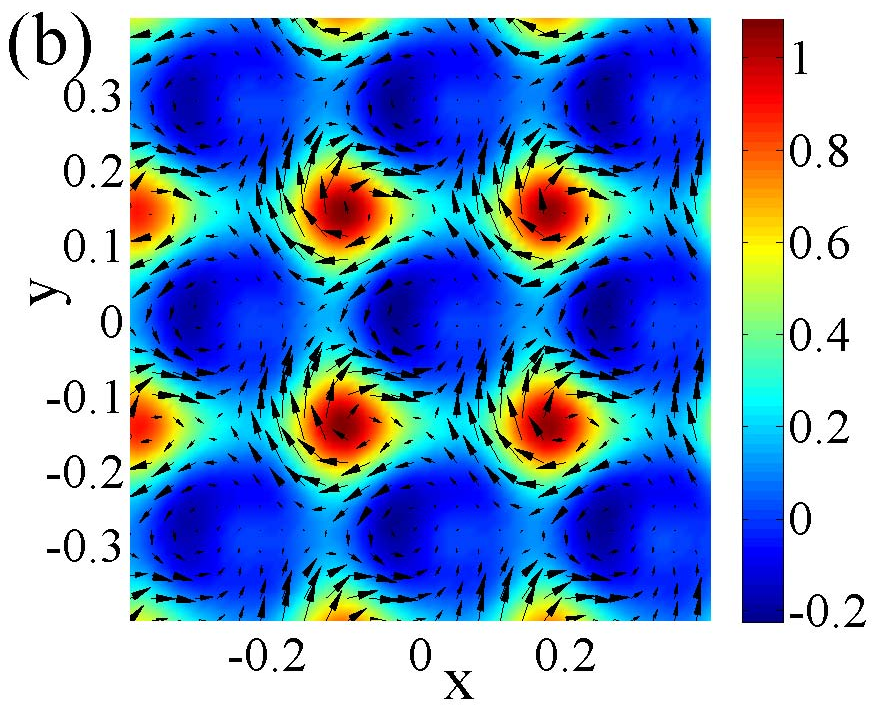}
\captionof{figure}{The distribution of $\vec S(\vec r)$
in horizontal cross-sections with (a) $z/l_T=0$, (b) $z/l_T=0.2$,
respectively.
The color scale shows the value of $S_z$
and parameter values are $\alpha=22$, $\beta=1$, and $c=1$.
The overall lattice exhibits the bcc structure.
}
\label{fig:lattice}
\end{center}
\end{widetext}
%--------------------------------------------------------

%%%%%%%%%%%%%%%%%%%%%%%%%%%%%%%%%%%%%%%%%%%%%%%%%%%
\section{The effect of strong interactions}
\label{sect:strong}

In this section, we present the condensate configurations
in the case that both spin-orbit coupling and interactions are
strong, such that different Landau levels are mixed by interactions.

In this case, the effect of the harmonic trapping potential becomes 
weak compared with interaction energies, thus we can approximate
the condensate wavefunctions as superpositions of plane-wave states. 
The plane-wave components are located on the
spin-orbit sphere and the condensate wavefunctions are no longer topological.
At $c=1$, the interaction is spin-independent, and bosons select a
superposition of a pair of states $\pm \vec k$
on the spin-orbit sphere, say, $\pm k_{so} \hat z$.
The condensate wavefunction is written as
\bea
\psi(\vec r)= \sqrt {\frac{N_a}{N_0}} e^{ik_{so}z} |\uparrow\rangle
+\sqrt {\frac{N_b}{N_0}} e^{-ik_{so}z} |\downarrow\rangle,
\label{eq:plane-wave}
\eea
with $N_a+N_b=N_0$.
The density of Eq. \ref{eq:plane-wave} in real space is uniform to
minimize the interaction energy at the Hartree-Fock level.
However, all the different partitions of $N_{a,b}$ yield the same
Hartree-Fock energy.
The quantum zero point energy from the Bogoliubov modes removes
this accidental degeneracy through the ``order-from-disorder'' 
mechanism, which selects the equal partition $N_a=N_b$.
The calculation is in parallel to that of the 2D Rashba case performed
in Ref. [\onlinecite{wu2008}], thus will not be presented here.
In this case, the condensate is a spin helix propagates
along $z$-axis and spin spirals in the $xy$-plane.

At $c\neq 1$, the spin-dependent part of the interaction can be written
as
\bea
H_{sp}=\frac{1-c}{2}g \int d^3 r
(\psi^\dagger_\uparrow \psi_\uparrow -\psi^\dagger_\downarrow \psi_\downarrow)^2.
\eea
At $c>1$, the interaction energy at Hartree-Fock level is minimized
for the condensate wavefunction of a plane wave state
$e^{ik_{so} z}|\uparrow\rangle$, or, its TR partner.

For $c<1$, $\langle H_{sp} \rangle$ is minimized if $\avg{S_z}=0$ in space.
At the Hartree-Fock level, the condensate can either be a plane-wave state
with momentum lying in the equator of the spin-orbit sphere 
and spin polarizing in the $xy$-plane,
or, the spin spiral state described by Eq. \ref{eq:plane-wave} with $N_a=N_b$.
An ``order-from-disorder'' analysis on the Bogoliubov zero-point energies
indicates that the spin spiral state is selected.
We also  present the numerical results for Eq. (4) in the main text
with a harmonic trap in Fig. \ref{fig:plane-wave} 
for the case of $c<1$.
The condensate momenta of two spin components have opposite signs,
thus the trap inhomogeneity already prefers the spin spiral state
Eq. \ref{eq:plane-wave} 
at the Hartree-Fock level.

%%%%%%%%%%%%%%%%%%%%%%%%%%%%%%%%%%%%%%%%%%%%%%%%%%%%%%%%%%%%%%
\begin{widetext}
\begin{center}
\includegraphics[width=0.3\linewidth]{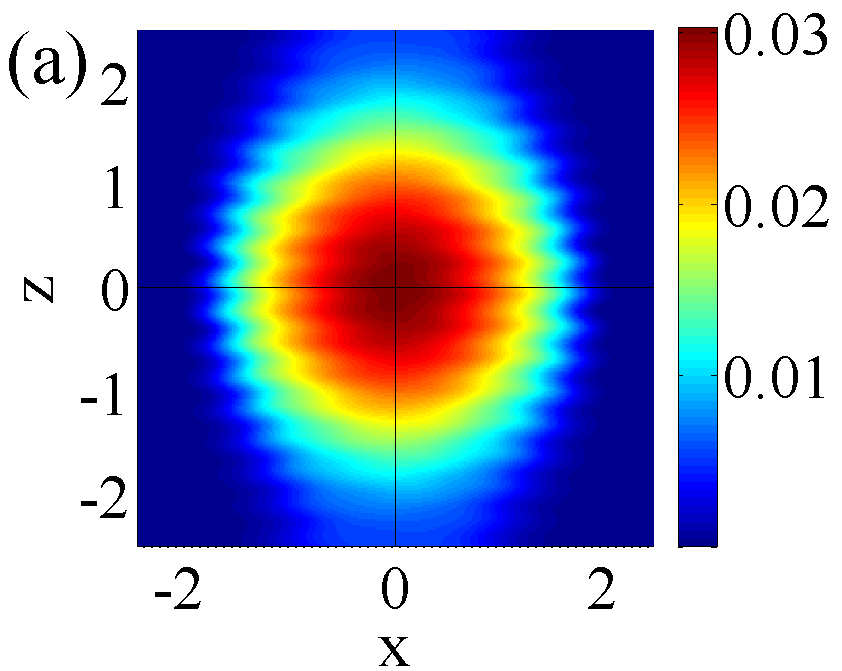}
\includegraphics[width=0.3\linewidth]{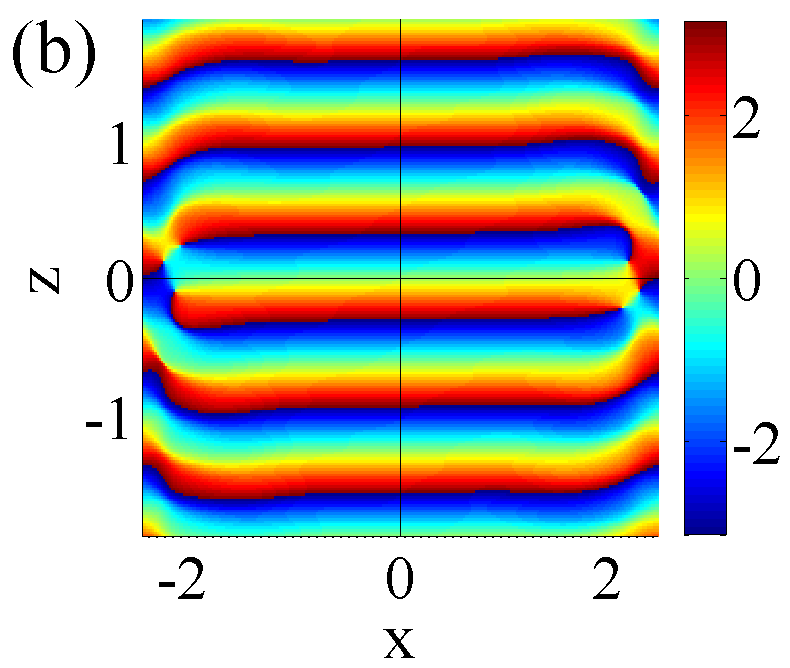}
\includegraphics[width=0.3\linewidth]{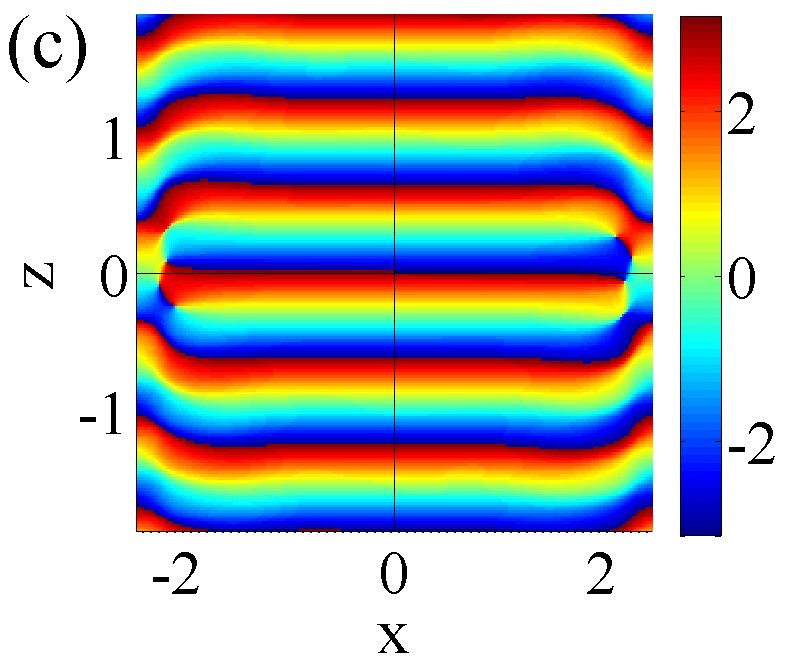}
\captionof{figure}{The density profile (a) for $\uparrow$-component,
and that for $\downarrow$-component is the same.
Phase profiles for (b) $\uparrow$ and (c) $\downarrow$-
components, respectively.
Parameter values are $a=10$, $\beta=50$, and $c=0.5$.}
\label{fig:plane-wave}
\end{center}
\end{widetext}
%%%%%%%%%%%%%%%%%%%%%%%%%%%%%%%%%%%%%%%%%%%%%%%%%%%%%%%%%%%%%%%%%%%%%%

\section{Conclusion}
\label{sect:conc}
In summary,  we have investigated the two-component unconventional BECs
driven by the 3D spin-orbit coupling.
In the quaternionic representation, the quaternionic phase distributions
exhibit non-trivial 3D skyrmion configurations from $R^3$ to $S^3$.
The spin orientation distributions exhibit texture configurations
characterized by non-zero Hopf invariants from $R^3$ to $S^2$.
These two topological structures
are connected through the 1st Hopf map from $S^3$ to $S^2$.
At large spin-orbit coupling strength, the crystalline order of spin textures,
or, wavefunction skyrmions, are formed, which can be viewed as
a generalization of the Abrikosov lattice in 3D.

{\it Note added.---} Near the completion of this manuscript, we became aware
of a related work by Kawakami {\it et al.} \cite{kawakami2012}, in which
the condensate wavefunction in the weak spin-orbit coupling case was studied.

{\it Acknowledgments.---} 
Y.L. thanks the Princeton Center for Theoretical Science at 
Princeton University for support.
X.  F.  Z.  acknowledges the  support of  NFRP  (2011CB921204,  
2011CBA00200),  the Strategic Priority Research Program of the 
Chinese Academy of Sciences (Grant No. XDB01030000), NSFC 
(11004186,11474266), and the Major Research plan of the National 
Natural Science Foundation of China (91536219).
C. W. is supported by the NSF DMR-1410375 and AFOSR FA9550-14-1-0168. 
C. W. acknowledges the support from the Presidents Research Catalyst
Awards of University of California, and National Natural Science 
Foundation of China (11328403).

\end{document}